\documentclass[%
 reprint,
 superscriptaddress,
 amsmath,amssymb,
 aps,
 prl,
 longbibliography,
lengthcheck,%
]{revtex4-1}

\usepackage{graphicx}% Include figure files
\usepackage{dcolumn}% Align table columns on decimal point
\usepackage{bm}% bold math
\usepackage{color}
\usepackage{ulem}

\begin{document}

\preprint{APS/123-QED}

\title{
Emergent Electromagnetic Induction and Adiabatic Charge Pumping in Weyl Semimetals
}

\author{Hiroaki Ishizuka}
\affiliation{
Department of Applied Physics, The University of Tokyo, Bunkyo, Tokyo, 113-8656, JAPAN 
}

\author{Tomoya Hayata}
\affiliation{
Department of Physics, Chuo University, 1-13-27 Kasuga, Bunkyo, Tokyo, 112-8551, JAPAN 
}

\author{Masahito Ueda}
\affiliation{
Department of Physics, The University of Tokyo, 7-3-1 Hongo, Bunkyo, Tokyo, 113-8656, JAPAN 
}
\affiliation{
RIKEN Center for Emergent Matter Sciences (CEMS), Wako, Saitama, 351-0198, JAPAN
}

\author{Naoto Nagaosa}
\affiliation{
Department of Applied Physics, The University of Tokyo, Bunkyo, Tokyo, 113-8656, JAPAN 
}
\affiliation{
RIKEN Center for Emergent Matter Sciences (CEMS), Wako, Saitama, 351-0198, JAPAN
}

\date{\today}

\begin{abstract}
The photovoltaic effect in a Weyl semimetal due to the adiabatic quantum phase is studied. We particularly focus on the case in which an external ac electric field is applied to the semimetal. In this setup, we show that a photocurrent is induced by the ac electric field. By considering a generalized Weyl Hamiltonian with nonlinear terms, it is shown that the photocurrent is induced by circularly, rather than linearly, polarized light. This photovoltaic current can be understood as an emergent electromagnetic induction in the momentum space; the Weyl node is a magnetic monopole in the momentum space, of which the electric field is induced by the circular motion. This result is distinct from conventional photovoltaic effects, and potentially useful for experimentally identifying Weyl semimetals in chiral crystals.
\end{abstract}

\pacs{
}% PACS, the Physics and Astronomy
% Classification Scheme.

\maketitle

{\it Introduction} --- The non-trivial phase in quantum adiabatic processes -- Berry's phase -- is one of the fundamental aspects of quantum mechanics. In a quantum system, the presence of an energy gap often prohibits excitation to higher-energy states, and confines electrons within a subspace of the Hilbert space constituted from lower energy states. In dynamical processes, such confinement sometimes gives rise to an additional geometric phase that depends only on the path, not on the details of dynamics.

Ever since its first discovery~\cite{Berry1984}, it has been revealed that Berry's phase leads to rich physics distinct from classical systems. Interestingly, the effect of Berry's phase appears not only in mesoscopic systems, but also in macroscopic properties of bulk materials. In solid-state materials, Berry's phase of electrons leads to non-trivial properties of solids, such as fractional pseudorotation quantum numbers in Jahn-Teller systems~\cite{Herzberg1963,Longuet-Higgins1975,Sakurai1993} and topological Hall effects~\cite{Ohgushi2000,Taguchi2001} arising from non-collinear magnetic textures.  A similar non-trivial structure of wave functions shows up in the Brillouin zone, and contributes to non-trivial structures in electronic states~\cite{Thouless1982,Kane2005}, and to transport phenomena~\cite{Sundaram1999,Moore2010,Sodemann2015}. 

Berry's phase also affects the dynamics of non-equilibrium systems. In periodically driven systems, it is known that the adiabatic phase induces the quantized pumping of charge~\cite{Thouless1983,Niu1984,Xiao:2009rm}. In an insulator, the pumping of charge is related to the time average of the emergent electric field defined by~\cite{Thouless1983}
\begin{eqnarray}
e^a(\vec{k}) = \partial_t A^a(\vec{k}) - \partial_a A^t(\vec{k}),\label{eq:emE}
\end{eqnarray}
where 
\begin{eqnarray}
A^a(\vec{k}) = -i\left<u_{n\vec{k}}(t)\right|\partial_a\left| u_{n\vec{k}}(t)\right>
\end{eqnarray}
is Berry's connection in the momentum space with $\partial_a=\partial/\partial k_a$ \ ($a=x,y,z$) and $\partial_t=\partial/\partial t$. However, in solids, the contribution from such an effect is usually zero or vanishingly small in a realistic setup, where the energy scale of the driving fields is much smaller than that of the band width.

\begin{figure}
  \includegraphics[width=\linewidth]{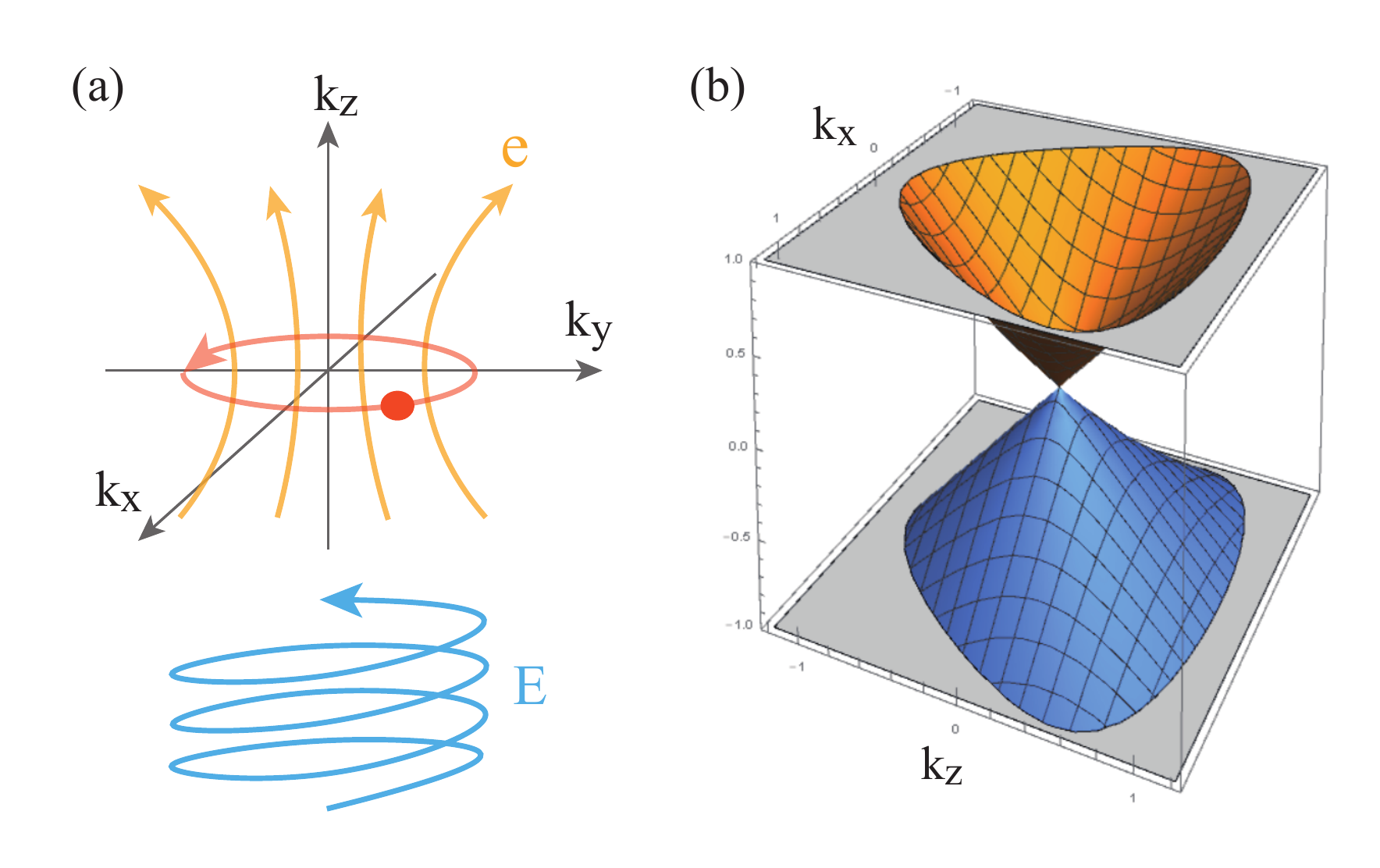}
  \caption{(Color online) (a) Schematic figure of the emergent electric field induced by an incident light. The red dot indicates the position of a Weyl node in the presence of an incident light. The incident electric field induces a rotational motion of the Weyl node through a Rashba-like coupling. The orbital motion of the Weyl node induces a dc emergent electric field penetrating through the orbit (yellow lines), that is, parallel to the propagation direction of the light. (b) Dispersion relations of the Weyl Hamiltonian with quadratic terms, $v=v_z=1$, $\alpha_1=0$ and $\alpha_2=-1$.}
  \label{fig:weyl_node}
\end{figure}
In this Letter, we theoretically show that such an effect of $\vec{e}$ field to the charge pumping is enhanced in Weyl semimetals, and possibly leads to experimentally observable consequences. Intuitively, this could be understood from the interpretation of Weyl nodes as the ``magnetic monopoles'' of Berry's connection. Suppose that we have a Weyl node at a nonzero $k_z$ and $k_x=k_y=0$. In such a situation, a Rashba-like coupling of electrons to the external electric field $\vec{E}$ induces a shift of the node in the $k_x$-$k_y$ plane. In the case of a circularly polarized light, the incident light results in a rotational motion of the Weyl node as schematically shown in Fig.~\ref{fig:weyl_node}(a). In analogy to the symmetric Maxwell's equation in real space, the circular motion of a magnetic charge in the momentum space induces the dc $\vec{e}$ field penetrating through the orbit. Since the $\vec{e}$ field is related to the electric current, $\vec{j}\propto\vec{e}$~\cite{Thouless1983}, the incident light can induce a dc current along the $z$ axis. This is the emergent electromagnetic induction in the momentum space. 

As is shown later, the photocurrent arises only with the circularly polarized light, in contrast to the conventional anomalous photocurrents~\cite{Fridkin2001} and those induced through Berry's curvature~\cite{Moore2010,Sodemann2015}. Also, it does not require a change in the charge distribution function; this is another distinct feature from the photocurrents induced through Berry's curvature and that in Weyl semimetals with broken time-reversal symmetry~\cite{Taguchi2016,Ebihara2016}, in which a change in  the charge distribution function is necessary to induce a  photocurrent. We note that our ``intrinsic'' photocurrent is sensitive to the direction of the incident light, in contrast to the other photocurrents in Weyl semimetals. We also discuss that, experimentally, a good candidate to observe the photovoltaic effect is a Weyl semimetal with broken spatial-inversion symmetry, such as TaAs~\cite{Weng2015,Lv2015,Sun2015}.

In the following, we first elaborate on some general requirements for free fermion systems to have a net $\vec{e}$ field. Based on this picture, we discuss a potential enhancement of the $\vec{e}$ field in Weyl semimetals. For this purpose, in the latter half, we particularly consider a generalized Weyl Hamiltonian with nonlinear terms, which couples to an external electric field by a Rashba-like coupling. In this model, we show that shining the circularly polarized light induces a electric current parallel to $\vec{k}^{(i)}$, the vector connecting the $\Gamma$ point and the $i$th Weyl node. 

{\it $\vec{e}$ field in periodically driven systems} --- In a periodically driven system, where the Hamiltonian $H(\{D_i\})$ is driven by slowly varying parameters, $D_i=D_i(t)=D_i(t+T)$ ($i=1,2,\cdots,m$), the average $\vec{e}$ field over a period reads
\begin{eqnarray}
e^a_n(\vec{k})&=& \partial_a\int_0^T [-A^t_n(\vec{k})] dt'\label{eq:netE1}\\
              &=& \sum_{i\neq j} i\partial_a\int_S  dD_i\wedge dD_j \partial_{D_j}\left<u_{n\vec{k}}(t)\right|\partial_{D_i}\left| u_{n\vec{k}}(t)\right>, \nonumber\\\label{eq:netE2}
\end{eqnarray}
where the region $S$ indicates a surface enclosed by the path given by $\{D_i(t)\}$ ($0\le t < T$). This is an $m-1$ dimensional hypersurface, and the integral does not depend on the choice of the surface. The integral in Eq.~(\ref{eq:netE1}) is Berry's phase in an adiabatic process~\cite{Berry1984}. From Eq.~(\ref{eq:netE2}), if $m=1$, there is no net $\vec{e}$ field in periodically driven systems because the area covered by the integral in Eq.~(\ref{eq:netE2}) is zero.

For a fully filled electron band, the average $\vec{e}$ field over a period of cycle is given by
\begin{eqnarray}
e^a(\vec{k})=- \partial_a\int_0^T \!\! A^t(\vec{k}) dt' ,
\end{eqnarray}
and the sum over the Brillouin zone reads
\begin{eqnarray}
  \bar{e}^a &=& -\int \prod_{b\ne a} dk_b \left.\int_0^T \! \! \partial_a A^t(\vec{k}) dt'\right|^{\pi}_{-\pi} .
\label{eq:thouless}
\end{eqnarray}
Here, we set the lattice constant to unity. It has been pointed out that, in insulators, the charge pumped during the adiabatic process is proportional to Eq.~\eqref{eq:thouless}~\cite{Thouless1983}, and that the integrand in the right-hand side of Eq.~\eqref{eq:thouless} gives a quantized value due to the single-valueness of the wavefunction.

In an insulator, however, it is expected that $e^a$ induced by an external electromagnetic field generally remains zero since the pumped current is a topologically protected quantity, and the external field is perturbatively small. In a two-band model, this can be seen from the fact that $\bar{e}^a$ is given by
\begin{eqnarray}
  \bar{e}^a=\frac12\int d^3k\int_0^T \! \! dt\, \vec{\hat{R}}(\vec{k},t)\cdot\partial_a\vec{\hat{R}}(\vec{k},t)\times\partial_t\vec{\hat{R}}(\vec{k},t).\nonumber\\
  \label{eq:2band}
\end{eqnarray}
Here, $\vec{\hat{R}}(\vec{k},t)$ is the normalized vector of $R_\nu(\vec{k},t)$ ($\nu=x,y,z$), and $\hat{R}_\nu(\vec{k},t)=R_\nu(\vec{k},t)/R$ with $R=|\vec{R}(\vec{k},t)|$. The Hamiltonian is given by
\begin{eqnarray}
  H(\vec{k},t)=\sum_\nu \sigma_\nu R_\nu(\vec{k},t),
\end{eqnarray}
where $\sigma_\nu$ ($\nu=x,y,z$) are Pauli matrices. The right-hand side of Eq.~(\ref{eq:2band}) gives the number of times $\vec{\hat{R}}$ wraps an unit sphere upon mapping $(k_\mu,t)\to S^2$ by $\vec{R}(\vec{k},t)$. In an insulator, since the energy scale of an external field is typically much smaller than that of electron band width, we naturally expect that this wrapping number becomes zero. 

In a doped case, the contribution from doped carriers gives a nonzero $e^a_n$. In a slightly doped insulator, however, such a contribution remains very small since the surface of the sphere covered in Eq.~\eqref{eq:2band} remains very small.

{\it Nonlinear Weyl Hamiltonian} --- An exception to such cases, in which charge doping leads to a large $e^a_n$ field, is a Weyl semimetal. In a Weyl semimetal, the effective Hamiltonian at the node is given by $H=0$. Hence, the Hamiltonian close to the node is dominated by external fields. As a consequence, a large $e^a_n$ field is expected by doping carriers to the node. To study the $\vec{e}$ field in periodically driven systems, we consider a doped Weyl Hamiltonian with nonlinear terms and multiple external fields:
\begin{subequations}
  \begin{align}
    R_x(\vec{k})&= v k_x + g D_y + \frac{\alpha_2}2 k_xk_z, \\
    R_y(\vec{k})&= v k_y - g D_x + \frac{\alpha_2}2 k_yk_z, \\
    R_z(\vec{k})&= v_z k_z + \frac{\alpha_1}2 (k_x^2+k_y^2-2k_z^2),
  \end{align}\label{eq:Hweyl}
\end{subequations}
where $k_a$'s are wave numbers at the Weyl node. We take here the local axis of $k$ such that the pair of nodes connected by time-reversal or spatial-inversion symmetry are along the $z$ axis, i.e., $\vec{k}^{(i)}$ is parallel to the $z$ axis. In Eq.~(\ref{eq:Hweyl}), the second term in $R_x(\vec{k})$ and $R_y(\vec{k})$ are the couplings with an external electric field of frequency $\omega$ and phase shift $\chi$:
\begin{eqnarray}
  D_x&=&D\cos(\omega t),\\
  D_y&=&D\sin(\omega t+\chi).
\end{eqnarray}
The electric field is circularly polarized for $\chi=0,\pi$, while it is linearly polarized for $\chi=\pi/2, 3\pi/2$. These Rashba-like couplings are allowed in general, if Weyl nodes are located away from symmetric points. When Weyl nodes are close to $\Gamma$ points, these terms appear from a coupling like
\begin{eqnarray}
  H_{el}=\tilde{g} \epsilon_{abc}D_a \kappa_b \hat{O}_c,\label{eq:rashba}
\end{eqnarray}
where $\kappa_b$ is the wave number from the $\Gamma$ point, $\hat{O}_\alpha$ ($\alpha=x,y,z$) is a set of operators that transform as vectors, and $\epsilon_{abc}$ is the Levi-Civita symbol. The third (second) term in $R_x(\vec{k})$ and $R_y(\vec{k})$ [$R_z(\vec{k})$] are quadratic in $k_a$. These terms break the $C_2$ rotation about an axis in the $xy$ plane, i.e., $+k_z$ and $-k_z$ become asymmetric as shown in Fig.~\ref{fig:weyl_node}(b). These terms reflect the existence of the pair node. Since our Hamiltonian includes information of the presence of the pair node, which always exists in a material, we believe that our Hamiltonian in Eq.~(\ref{eq:Hweyl}) is a generic model for Weyl semimetals in solids.

We first consider the case of electron doping. To evaluate the $\vec{e}$ field, we focus on the limit $\alpha_i\ll v/|\mu|, v_z/|\mu|$ where $\mu$ is the chemical potential ($\mu>0$ and $\mu<0$ for electron and hole doping, respectively). To calculate $\bar{e}_n^a$, we expand the $\vec{e}$ field up to second order in $\alpha_i$.

The filling of electrons are fixed at $D_x=D_y=0$. To take into account the change of the Fermi surface by $\alpha_i$, we expand the dispersion relation around the Fermi surface for $\alpha_1=\alpha_2=0$ along the radial direction. For the electron doped case, the change in $k_F$, $\Delta k$, can be calculated by solving 
\begin{eqnarray}
  \mu-\varepsilon(\vec{k}_F^{(0)})+\Delta k\partial_k \varepsilon(\vec{k}_F^{(0)})=0,
\end{eqnarray}
where $\vec{k}_F^{(0)}$ is the Fermi surface for $\alpha_1=\alpha_2=0$. From the fact that $\mu-\varepsilon(\vec{k}_F)\sim O(\alpha_i)$, we expect $\Delta k\sim O(\alpha_i)$. Hence, in general, we need to consider terms up to $O(\Delta k^2)$ to fully take into account terms up to $O(\alpha_i^2)$. However, from explicit calculation, we find that the $O(\alpha_i^2)$ contribution to $\Delta k$ vanishes.

The second order response in $D$, up to $O(\alpha_i^2)$, gives a net emergent electric field along the $z$ axis:
\begin{eqnarray}
  \bar{e}^z_\pm = \pm\pi\frac{4(v^2-2v_z^2)\alpha_1-3vv_z\alpha_2}{30v^5v_z^3}\alpha_1 (\mu g D)^2\omega\cos(\chi),\nonumber\\\label{eq:Eweyl}
\end{eqnarray}
for Weyl ($+$) and anti-Weyl nodes ($-$), respectively. Due to the phase factor $\cos(\chi)$, the $\vec{e}$ field shows a maximum for a circular light ($\chi=0,\pi$), while it vanishes for a linearly polarized light ($\chi=\pi/2,3\pi/2$). This is consistent with the general argument above, and indicates the absence of a dc $\vec{e}$ field when we have only one time-dependent parameter.

For electron doping, the net emergent electric field increases as a function of $\mu^2$. This indicates that the contribution from electrons with energy $\varepsilon$ decays like $\vec{e}\sim \varepsilon^{-1}$, since the density of states is approximately proportional to $\varepsilon^2$. This implies that, when considering hole doping, we need to appropriately take into account the contribution from electron states in the UV limits. However, from the argument above, $\bar{e}$ naturally vanishes for the filled bands. Therefore we can evaluate the $\vec{e}$ field for the hole doped case by subtracting the contribution from vacant states. For the model in Eq.~(\ref{eq:Hweyl}), the result becomes the same as  Eq.~(\ref{eq:Eweyl}) with flipping the overall sign.

{\it Discussions} --- In the mechanism presented here, the coupling of the external electric field to the electron orbitals, e.g., the Rashba-like coupling, plays an important role. This is due to the fact that Berry's phase arises from the non-trivial change of the Bloch wave function over the period of the cycle. It gives rise to a different consequence from the Peierls substitution terms, of which the nonlinear responses have been studied recently~\cite{Chan2016,Taguchi2016,Ebihara2016}.

In our result, the photocurrent is induced by the adiabatic dynamics of electron orbitals; a change in the electron distribution function is not required. Also, since the current is proportional to Berry's phase, the photocurrent arises only for the circularly polarized light while no current arises for the linearly polarized light. Another important feature of the coupling to electron orbitals is its anisotropy. Since Weyl nodes in solids are generally located away from symmetric points in the Brillouin zone, the coupling to electron orbitals is generally anisotropic. For instance, in the case of the coupling given by Eq.~(\ref{eq:rashba}), in the lowest-order approximation, the coupling exists only for the $x$ and $y$ directions as in the Hamiltonian in Eq.~(\ref{eq:Hweyl}). As a consequence, the photocurrent is expected to be highly sensitive to the direction of the incident light.  
%This is a distinct feature from the photovoltaic effects in the inversion symmetry broken materials~\cite{Fridkin2001}, where the electric current is induced by a linearly polarized light. It is also distinct from the photocurrent from the Berry curvature in $\vec{k}$ space~\cite{Moore2010,Sodemann2015}. For these photovoltaic effects, change in the Fermi distribution function is crucial. Also, the current can be induced by a linearly polarized light. In our setup, in contrast, no photocurrent is induced for the linearly polarized light. This is due to the fact that the current is induced from the adiabatic Berry's phase, which requires the existence of multiple time-dependent parameters.
%Recently, photovoltaic effects in Weyl semimetals with a similar setup have been proposed~\cite{Ebihara2016,Taguchi2016}.

In addition to the orbital coupling terms, the Peierls substitution terms can contribute to photocurrents in the Weyl semimetals. Such photovoltaic effects in Weyl semimetals have been theoretically studied in the case of broken time-reversal symmetry~\cite{Taguchi2016,Ebihara2016}. In these theories, however, the change in the electron distribution function, either by chiral magnetic effect~\cite{Ebihara2016} or by transferring the photon angular momentum to electrons~\cite{Taguchi2016}, plays a key role in the photovoltaic effect. In contrast, the mechanism proposed in this Letter does not involve the change in the electron distribution function. Also, the photocurrent from Peierls substitution terms is expected to be isotropic since the Hamiltonian in the lowest order approximation preserves $SO(3)$ rotational symmetry.

\begin{figure}
  \includegraphics[width=\linewidth]{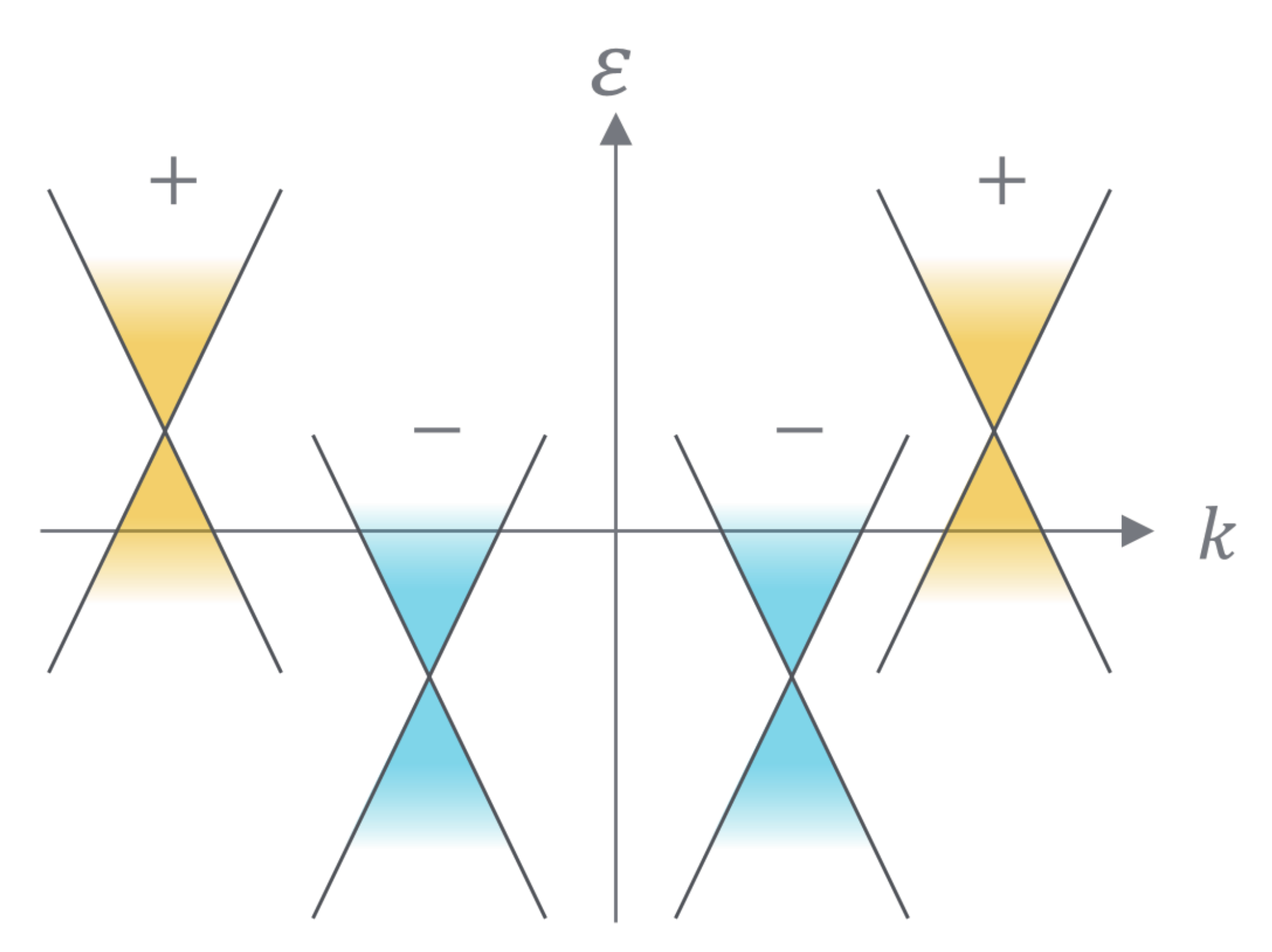}
  \caption{(Color online) Schematic picture of a Weyl semimetal with broken spatial-inversion symmetry. Each cone indicates a Weyl node and the sign shows the chirality. In the presence of time-reversal symmetry, two Weyl nodes with the same chirality are related by the symmetry operation, so that there are at least four Weyl nodes in the Brillouin zone.}
  \label{fig:weylsm}
\end{figure}

In regard to materials, recently, Weyl semimetals in solids has been explored in various materials~\cite{Wan2011,Burkov2011,Xu2011,Fang2012,Chen2013,Yamaji2014,Guan2015,Ueda2015,Tian2016,Huang2016,Chen2016,Weng2015,Lv2015,Sun2015}. From symmetry argument, in solids, the presence of Weyl nodes requires breaking of either time-reversal or spatial-inversion symmetry. In the case of Weyl semimetals with broken time-reversal symmetry, and in the presence of spatial-inversion symmetry, a Weyl node has a pair anti-Weyl node. Since these two nodes are related by spatial-inversion symmetry, when a Weyl node is doped, there always exists an anti-Weyl node with exactly the same doping. For the photocurrent, since the sign of induced current depends on the chirality, the effect of electric field always cancels out, and total $\vec{e}$ field (electric current) becomes zero. This is consistent with the fact that a photovoltaic effect generally requires the breaking of spatial-inversion symmetry.

In contrast, in the case of Weyl semimetals with broken spatial-inversion symmetry, a Weyl (an anti-Weyl) node has a pair Weyl (a pair anti-Weyl) node~\cite{Weng2015,Lv2015,Sun2015}. Hence, there always exist at least four nodes in the Brillouin zone (see Fig.~\ref{fig:weylsm}). In such materials, the parameters in the Hamiltonian in Eq.~(\ref{eq:Hweyl}) are different between pairs of Weyl nodes. The doping level ($\mu$) also differs for each pair. Therefore, in this setup, the emergent electric fields from Weyl and anti-Weyl pairs have different values, so that the net $\vec{e}$ field becomes non-zero. We emphasize here that Weyl semimetals with broken spatial-inversion symmetry are preferrable for observation of the transport phenomena induced by emergent electric fields.

We also note that even in the Weyls semimetals with broken time-reversal symmetry, it might be possible by the chiral magnetic effect~\cite{Nielsen:1983rb,Fukushima2008,Son2013,Hayata2016} to make a difference between doping levels of Weyl/anti-Weyl pairs, and to break the cancelation of photocurrent in them. In this mechanism, application of dc electric and magnetic fields induces chiral charge proportional to the inner product of electric and magnetic fields. Hence, consideration of the chiral magnetic effect leads to a non-zero photocurrent. In this case, the photocurrent is observed as a nonlinear part of the conductivity, where  its sign changes by changing the polarization from the right to the left hand.

Besides the Weyl semimetals, the general argument on the enhancement of the $\vec{e}$ field can be applied to other nodal (semi)metals, such as the surface state of topological insulators, double Weyl~\cite{Huang2016,Chen2016} and Dirac~\cite{Hsieh2008,Li2008} semimetals as well as those with quadratic band touching~\cite{Sun2009,WitczakKrempa2012}. However, the coupling of nodes to external electric fields differs for these systems, and leads to quantitatively different consequences. Quantitative analysis of the $\vec{e}$ field in these systems is a important direction to engineer unconventional responses as well as to develop an experimental method to probe electronic structures of materials. It is left for future studies.

%{\it Summary} --- To summarize, in this paper, we have studied electric transport phenomena in Weyl nodes. By focusing on the small $\omega$ limit, we show that the electric transport from an ac field can be strongly enhanced in the case of Weyl semimetals. By considering a generalized Weyl Hamiltonian with $O(k^2)$ terms, we have studied electric transport in the doped Weyl semimetals. By applying electric field, we show that the circularly polarized light gives rise to a dc electric current along $\vec{k}^{(i)}$, the vector connecting $\Gamma$ point and the Weyl node; the dc current only appears for circular light, and is absent for linearly polarized light. In materials, these phenomena are expected to be observed in Weyl semimetals with broken spatial-inversion symmetry. As this photovoltaic effect have different features from conventional ones, observation of this phenomena may be useful as a method to experimentally identify Weyl semimetals.

The authors thank M. Ezawa and K. Kikutake for fruitful discussions. This work was supported by JSPS Grant-in-Aid for Scientific Research (No. 24224009, No. 26103006, No. 26287088, No. 15H05855, and No. JP16J02240), from MEXT, Japan, and ImPACT Program of Council for Science, Technology and Innovation (Cabinet office, Government of Japan).

\end{document}